\documentclass[10pt]{article}
\usepackage{fullpage,citesort,epsfig,graphics,amsbsy}
\usepackage{psfrag}
\usepackage[small]{caption}
\usepackage{Macro_Thorn}
\usepackage{Macro_Qiu}
\usepackage{slashed}
\newcommand \dreg{HV}
\newcommand \dred{DR}
\newcommand {\glug} {g-}
\newcommand {\gluG} {g+}
\newcommand {\fl}[1] {q_{#1}-}
\newcommand {\fr}[1] {q_{#1}+}
\newcommand {\Fl}[1] {\bar{q}_{#1}-}
\newcommand {\Fr}[1] {\bar{q}_{#1}+}
\newcommand {\pref} {-2ig^2\trace(t^at^bt^ct^d)}
\begin{document}
\setlength{\captionmargin}{20pt}
\begin{titlepage}
\begin{flushright}
UUITP-03-08\\
\end{flushright}

\vskip 3cm

\begin{center}
\begin{Large}
{\bf Dimensional Regularization and Dimensional Reduction in the
Light Cone}
\end{Large}

\vskip 2cm {\large J.Qiu} \vskip0.20cm {Department of Physics and
Astronomy\\ Uppsala University\\ Sweden}

(\today)

\vskip 1.0cm
\end{center}

\begin{abstract}\noindent
We calculate all the 2 to 2 scattering process in Yang-Mills
theory in the Light Cone gauge, with the dimensional regulator as
the UV regulator. The IR is regulated with a cutoff in $q^+$. It
supplements our earlier work, where a Lorentz non-covariant
regulator was used and the final results bear some problems in
gauge fixing. Supersymmetry relations among various amplitudes are
checked using the light cone superfields.
\end{abstract}
\vfill
\end{titlepage}
\section{Introduction}%
The recent years saw a fast development in the techniques of
perturbative computation of amplitudes in Yang-Mills theory. Among
these are string based computation \cite{GreenSchwarzBrink},
unitarity based techniques \cite{unitarity}, and MHV techniques
\cite{CSW} inspired by the twistor string formalism. Thanks to
these new developments, a large class of amplitudes can now be
computed. In particular, for the N=4 supersymmetric Yang-Mills
(MSYM) theory, certain amplitudes have been calculated to quite
high loop orders \cite{highloops}. And an iteration relation to
all orders was postulated by Bern, Dixon and Smirnov (BDS)
\cite{BDS}. This iteration relation became all the more
interesting in the light of \cite{AldayMaldacena}, where the gluon
scattering amplitude is calculated at strong coupling using
AdS/CFT\footnote{a review paper \cite{LJDixon} become available
during the preparation of the paper}.

The present article is dedicated to the computation of the
scattering amplitudes in Yang-Mills theory in the light cone
gauge. The purpose of this work is, however, not to bring coals to
Newcastle. Admittedly, our ability to perform computation in the
light cone gauge is severely marginalized when compared to the
others using advanced and streamlined techniques. But there are
plenty of occasions when a 4D computation is desirable, or when we
would like each field in the lagrangian to correspond to one
particle exactly. The first of our series of computations was
necessitated by the need to study the renormalization of YM theory
represented on the light cone world-sheet (see \cite{I} for an
introduction), in the selfsame paper we found some novel (local
world sheet) counter terms required to restore gauge invariance.
Recently, when attention was paid to the search for the Lagrangian
origin of MHV rules \cite{LagOriMHV}, the {\small++}$-$ (non-MHV)
vertex present in the light cone lagrangian was absorbed into the
kinetic term through a change of variable, and one of the counter
terms found in \cite{I} was interpreted as part of the Jacobian
associated to the changing of variable \cite{BSTZ}. In addition to
the novel counter terms, we also found a nice infrared subtraction
scheme tailored for the light cone in \cite{II}. This nice
separation of an amplitude into a hard part(infrared free) and a
soft part was in fact a surprise to us, for neither party was
expected to be Lorentz covariant on its own with our simple
infrared subtraction.

In our past works \cite{I,II,III,IV}, we have computed the
scattering amplitudes in the light cone gauge with a
$\delta$-regulator (an exponential damping in the transverse
direction) to tame the UV. This regulator has some peculiar
features in terms of breaking the gauge invariance. Naturally all
the breaking of gauge invariance can be restored with proper
counter terms, but the little freedom in choosing the counter
terms leaves the amplitudes defined only up to a constant multiple
of a tree. Thus only in special cases such as MSYM, where all
counter terms vanish, can we check the supersymmetric relations
among various amplitudes. So we turn to using a dimensional
regulator specially modified for the light cone, first and
foremost as a check for the calculation done in our past work, and
yet we also expect it to have some spin off's. The main feature of
the dimensional regulator used in this article is that it only
regulates UV divergence leaving the IR regulated by setting the
$+$ component of the loop momenta away from zero as was done in
\cite{II}. This regularization scheme turned out just as
efficient, in terms that the final result is automatically gauge
invariant, and in the case of MSYM, bears no reference to the
dimension parameter $\epsilon$.

In this article, we use both the dimensional regularization or 't
Hooft-Veltman scheme \cite{tHooftVeltman}, and dimensional
reduction \cite{Siegel}. To save space we shall refer to the
former as HV scheme and the latter as DR scheme. HV scheme is
known to break supersymmetry
\cite{Siegel,CapperJonesNieuwenhuizen}, and does not go well with
the helicity based calculation. These are merely two different
facets of the same problem. In 4D, the orthochronous component of
$SO(3,1)$ has a double covering $SL(2,C)$, which enables us to
express a polarization vector as a product of two spinors of the
form $|\lambda]\bra{k}$, and by picking the reference spinor
$|\lambda]$ wisely, we spare ourself the trouble of computing a
large class of diagrams. Sometimes this makes all the difference
between being able or not able to perform certain computations.
One cannot use the same strategy in a different dimension, and a
related problem is whether or not the matrices of the form
$\gamma^{[\mu_1}\gamma^{\mu_2}...\gamma^{\mu_i]}$ form a complete
basis as they do in 4D. This would tell on the validity of Fierz
identity, and thereby the preservation of supersymmetry
\cite{BrinkSchwarzScherk}. With {\dred} scheme, all of the above
problems are absent, the result preserves gauge symmetry and
supersymmetry. Our modified {\dred} scheme produce results that
conform to this common lore.

This article is organized as follows, sec.\ref{tech_dreg} contains
the strategy of our dimensional regulation schemes and some
examples. Computational results are listed in
sec.\ref{amplitudes}. Although the loop calculations are done in
the component fields, we use light cone superfield formalism to
discuss the supersymmetric relations among various amplitudes in
sec.\ref{supersymmetry}. Sec.\ref{Infrared Terms} contains the
discussion of the infrared parts regulated in the light cone
fashion, which is the major novel feature of our regulation
scheme. In the appendix, some intermediate computational results
are listed and the light cone super field is also summarized
there.

\section{Dimensional Regularization in the Light Cone}\label{tech_dreg}
The light cone singles out the $t$ and $z$ direction of the space
time, leaving the $x$ and $y$ direction as the transverse
dimensions. So a natural choice of {\dreg} scheme suited for the
light cone is to set the number of transverse direction to be
$D=2-2\epsilon$. Now it is crucial that our dimensional regulator
does not touch upon the IR part, so any possible difference in the
result between using a UV $\delta$-regulator and dimensional
regulator will only come from the superficially UV divergent part
\footnote{Note the separation of an amplitude into UV divergent
part and finite part is highly gauge dependent, and even procedure
dependent, a point to bear in mind}.

\textit{With this observation, we need only focus on the
potentially divergent part of an amplitude, and study the possible
difference resulting in applying either $\delta$ or HV regulator.}

As stated in the introduction, a helicity based calculation is
much preferred, for it grants us the ability to apply some
on-shell identities of the amplitudes. We can apply this method
equally well in the light cone, with the polarization vectors
defined as\bea \epsilon_{\vee}^{\mu}=\frac{1}{\sqrt2}(\frac{k^1-
ik^2}{k^0+k^3},1,-i,-\frac{k^1-
ik^2}{k^0+k^3})=\frac{1}{\sqrt2}(\frac{k^{\wedge}}{k^+},1,-i,-\frac{k^{\wedge}}{k^+})\textrm{
; }\epsilon_{\wedge}=\epsilon_{\vee}^*\eea where the subscript
$\wedge$ $\vee$ correspond to the positive and negative helicity.
With the {\dreg} scheme, we have a total of $D=2-2\epsilon$
polarizations for a gluon. Then it become awkward to use the
$\wedge$ $\vee$ basis. In particular, we have problems
interpreting terms like $\delta^{\wedge}_i$, where $i$ ranges from
$2-2\epsilon$ to 2. Especially any error we make might be
amplified by a factor of $1/\epsilon$ and become non-negligible.

To avoid the ambiguities, we need to treat all the $2-2\epsilon$
transverse direction as equal by adopting a cartesian basis. In
the cartesian basis, the polarization vector relevant for the
light cone become\bea
\epsilon_i^\mu=(\frac{1}{\sqrt{2}}\frac{k^i}{k^+},...,\delta^{\mu}_i,...,-\frac{1}{\sqrt{2}}\frac{k^i}{k^+})\eea
where the subscript $i$ is a label that labels the '$i^{th}$'
polarity of a gluon. The $-$ component of $\epsilon_i^\mu$ is thus
$k^i/k^+$, the $+$ component is of course zero corresponding to
the light cone gauge fixing. While for the transverse components,
only the $i^{th}$ take a $1$ with all the others $0$. In the
appendix as well as in \cite{III}, we give some properties of this
polarization vector. With this setup, the transverse dimension are
treated on equal footing, instead of being further divided into a
2 and $D-2$ part, which will make the calculation rather untidy.
As for the amplitudes involving fermions, we face a larger problem
of manipulating the gamma matrices in a non-integer dimension, but
so long as we adopt the cartesian basis and stick to Dirac
fermions (staying away from $\gamma^5$), there is no conceptual
difficulty in the interpretation (as we shall see later).

We now give an example of this procedure. Consider a generic box
diagram, after the usual steps of exponentiating the denominators,
shifting the momenta etc, we can write down the divergent part of
this box integrand.
\bea B_{HV}=\mu^{2\epsilon}\int \frac{d^D
q_{\perp}}{(2\pi)^D}\int_0^\infty\frac{T^3dT}{2T}\int_{0<x_i<1}
d^4 x_i\delta(\sum {x_i}-1)
q_{\perp}^4\exp{\left(-iTq_{\perp}^2+iTH\right)}B(2+D)\eea where
$B(2+D)$ is an object that encapsules all the $D+2$ dimensional
tensor algebra or gamma matrix algebra and $H$ is just some
function of the Schwinger parameters that need not concern us now.
The same
integrand written with a $\delta$-regulator is%
\bea B_{\delta}=\int \frac{d^2
q_{\perp}}{(2\pi)^2}\int_0^{\infty}\frac{T^3dT}{2T} d^4
x_i\delta(\sum {x_i}-1)
q_{\perp}^4\exp{\left(-i(T-i\delta)q_{\perp}^2+iTH\right)}B(4)\eea%
The evaluation is straight forward, we get for {\dreg} scheme\bea
B_{HV}=\int d^4 x_i\delta(\sum {x_i}-1)\frac{i}{8\pi^2}
\frac{D(2+D)}{8}B(2+D)\left[\frac{1}{\epsilon}-\gamma\right]\left[1-\epsilon
\log{\frac{H}{\mu^2}}+\epsilon\log{4\pi}\right]\eea
where $\epsilon$ is $(2-D)/2$, and we can also redefine $\mu$ to
absorb $\log{4\pi}$. Using $\delta$-regulator, we have\bea
B_{\delta}=\int d^4 x_i\delta(\sum {x_i}-1)\frac{i}{8\pi^2}B(4)
\left[-\log{\delta H
e^\gamma}-\frac{3}{2}\right]\eea%
We are only interested in the difference between these two schemes
\bea B_{HV}-B_{\delta}=\int d^4 x_i\delta(\sum
{x_i}-1)\frac{i}{8\pi^2}\left[(\frac{1}{\epsilon}+\log{\mu^2})B(2+D)+\log{(\delta)}B(4)
\right]+\mathcal{O}(\epsilon)\eea
%\bea B_{HV}-B_{\delta}\sim \int d^4 x_i\delta(\sum
%{x_i}-1)\frac{i}{8\pi^2}\left[(\frac{1}{\epsilon}+\log{\mu^2})B(2+D)+\log{(\delta)}B(4)
%-(\log{He^{\gamma}}+\frac{3}{2})\left[B(2+D)-B(4)\right]\right]+\mathcal{O}(\epsilon)\eea
%
Clearly $1/\epsilon+\log{\mu^2}$ is to be identified with
$-\log{\delta}$, hence\bea
B_{HV}-B_{\delta}=\frac{i}{8\pi^2}\lim_{\epsilon\to
0}{\frac{1}{\epsilon}\left[B(2+D)-B(4)\right]}\label{ambi_problem}\eea
Note the dependence of $B(2+D)$ upon $D$ is two fold, one explicit
dependence for which we can differentiate, take limits etc; the
other intrinsic dependence that does not allow for such
manipulations. So some care is needed in dealing with
Eq.\ref{ambi_problem}, especially when gamma matrices are involved
(for example a term $\sum{\gamma^{i}\gamma^{i}}/\epsilon$, where
the summation is from $2-2\epsilon$ to 2, is hard to give meaning
to).

For self-mass and vertex corrections, we may try to compute the
wave function or charge renormalization (whose $D$ dependence is
obviously differentiable) to circumvent the possible ambiguity of
Eq.\ref{ambi_problem}. For the scattering amplitude, we will
compute the superficially divergent part, the terms that has
$1/\epsilon$ poles will be kept to the very end, where they have
to fall together to be proportional to a tree. Note this tree is
defined in $2+D$ dimension and this is done unambiguously only by
using cartesian basis. In the same manner for the
$\delta$-regulator, terms multiplying $\log{\delta}$ must also be
proportional to a tree (defined in $4$ dimension). At this stage
we will identify $(1/\epsilon+\log{\mu^2})\cdot
\textrm{tree}(2+D)$ with $-\log{\delta}\cdot \textrm{tree}(4)$.
After this identification, whatever is left will be finite but
will not in general be proportional to a tree, in fact, they will
have to compensate for the breaking of gauge invariance due to the
$\delta$-regulator (the example is given in sec.\ref{Assembling}).

\subsection{Box Diagrams}\label{boson_box}
We shall apply the above procedure to the box diagrams as follows:
denote by $B^{\mu\nu\rho\sigma}$ a generic four gluon box diagram,
with external indices unspecified. Let\bea
B(g^i,g^j,g^k,g^l)\equiv(-\epsilon_{i\mu})(-\epsilon_{j\nu})(-\epsilon_{k\rho})(-\epsilon_{l\sigma})B^{\mu\nu\rho\sigma}\eea
be a box where the four external gluons are physically polarized.
We emphasize that the indices $i,j...$ are not space time indices,
but rather labels that label the different polarities. Since a box
diagram in the light cone gauge is logarithmic divergent, the
indices $i,j...$ will only be carried by Kronecker delta symbols.
We get (the argument behind the semicolon:'g','s' or 'q' means
that gluon, scalar or fermion circulates the loop) \bea
B(g^i,g^j,g^k,g^l;g)&=&\frac{2}{2-D}\bigg\{\left[\delta^{il}\delta^{jk}+\delta^{ij}\delta^{kl}\right]\left[\frac{8}{3(2+D)}-\frac{2}{D}+\frac{1}{2}\right]+\delta^{ik}\delta^{jl}\frac{8}{3(2+D)}+\textrm{Art.
Div.}\bigg\}\nn\\
B(g^i,g^j,g^k,g^l;s)&=&\frac{2}{2-D}\left[\delta^{il}\delta^{jk}+\delta^{ij}\delta^{kl}+\delta^{ik}\delta^{jl}\right]\frac{8}{3D(2+D)}\nn\\
B(g^i,g^j,g^k,g^l;q)&=&\frac{2}{2-D}\bigg\{\left[\delta^{il}\delta^{jk}+\delta^{ij}\delta^{kl}\right]\frac{-4(3D^2-4)}{3D(2+D)}+\delta^{ik}\delta^{jl}\frac{4(3D^2-8+6D)}{3D(2+D)}\bigg\}\label{box_dreg}\eea
where we omit a factor $ig^4N_c/(8\pi^2)\trace{t^at^bt^ct^d}$. The
results are presented so that we can get the $1/\epsilon$ pole by
taking $D$ to $2$, or the difference HV versus $\delta$ by
extracting the finite part as $D\to 2$. Note that the difference
in the tensor structure among the above three is expected.
$B(g^i,g^j,g^k,g^l;q)$ must be finite when summed over all
possible cyclic orderings, corresponding to the finiteness of the
four photon box in QED. $B(g^i,g^j,g^k,g^l;g)$ has the same
property, which simply says photons do not interact with photons
in the absence of the electrons (called $U(1)$-decoupling).

'Art. Div' stands for 'gauge artificial divergence' and is poles
in $q^+$. They are too cumbersome to list and they in general will
be $D$ dependent. All the final results must go through the
non-trivial test that the gauge artificial divergences cancel up
to $\mathcal{O}(\epsilon)$.

Another example is $B(g^i,g^j,q,\bar{q})$, a box with two fermion
legs and two gluon legs. This diagram would have been finite by
naive power counting, however, in the light cone set up, we have
terms like $q^i/q^+$ that will be missed by the naive power
counting (we do not adopt the Mandelstam prescription, so a Wick
rotation is not valid), but actually contributing to the UV
divergence. But we can still see the trace of naive power counting
from the fact that its $1/\epsilon$ pole contribution consists
solely of gauge artificial divergences, i.e. of order
$\mathcal{O}(1/q^+)$, conforming to the result of naive power
counting.
\subsection{Self-mass Diagrams}
The bosonic self-mass diagrams are similar, and the results using
dimensional regularization must be gauge covariant(transverse). We
first list the $\delta$-regulator results (with the mass
renormalization term and a factor of
$ig^2N_c/(8\pi^2)\trace{[t^at^b]}$ omitted) \bea \Pi(g^i,g^j;s)
&=&\delta^{ij}p^2\left[\frac{5}{18}-\frac{1}{6}\log{p^2\delta
e^\gamma}\right]\hspace{1cm}\Pi(+,+;s)=p^{+2}\left[\frac{4}{9}-\frac{1}{6}\log{p^2\delta
e^\gamma}\right]\nn\\
\Pi(g^i,g^j;q)&=&\delta^{ij}p^2\left[\frac{26}{9}-\frac{4}{3}\log{p^2\delta
e^\gamma}\right]\hspace{1cm}\Pi(+,+;q)=p^{+2}\left[\frac{20}{9}-\frac{4}{3}\log{p^2\delta
e^\gamma}\right]\nn\\
\Pi(g^i,g^j,g)&=&\delta^{ij}p^2\left[-\frac{67}{9}+\frac{11}{3}\log{p^2\delta
e^\gamma}+\int_0^{p^+}{dq^+\left[\frac{-2}{p^+-q^+}+\frac{-2}{q^+}\right]\log{\frac{(p^+-q^+)q^+p^2\delta
e^\gamma}{p^{+2}}}}\right]\nn\\
\Pi(+,+,g)&=&p^{+2}\left[\frac{8}{9}-\frac{1}{3}\log{p^2\delta
e^\gamma}\right]\label{sf_mass_delta}\eea where $g^i$ tells the
polarity of the gluon, while $\Pi(+,+)$ describes the propagation
of $A^-$. These results are not Lorentz covariant, as can be seen
from the fact that the numbers in $\Pi(g^i,g^j)$ do not match
those in $\Pi(+,+)$. The procedure in the previous section must
produce a compensation to these mismatches. Indeed, when we
evaluate the superficially divergent parts of self-mass diagrams
in $2+D$ dimension\bea
\Pi(g^i,g^j;g)&=&\frac{2}{2-D}\delta^{ij}p^2(-\frac{11}{3})\hspace{1cm}%
\Pi(+,+;g)=\frac{2}{2-D}\frac{D}{6}p^{+2}\nn\\
\Pi(g^i,g^j;s)&=&\frac{2}{2-D}\delta^{ij}p^2(\frac{1}{3D})\hspace{1cm}%
\Pi(+,+;s)=\frac{2}{2-D}\frac{1}{6}p^{+2}\nn\\
\Pi(g^i,g^j;q)&=&\frac{2}{2-D}\delta^{ij}p^2(-\frac{4}{3D}+2)\hspace{1cm}%
\Pi(+,+;q)=\frac{2}{2-D}\frac{4}{3}p^{+2}\label{sf_mass_dreg}\eea
One can extract the finite terms as $D\to 2$ according to
Eq.\ref{ambi_problem} and combine them to Eq.\ref{sf_mass_delta},
and we gladly see that the Lorentz covariance is restored in
$\Pi(g,g;s)$ and $\Pi(g,g;q)$.
\bea\Pi(g,g;q)=-(p^2g^{\mu\nu}-p^\mu
p^\nu)\left[\frac{20}{9}-\frac{4}{3}\log{p^2\delta
e^\gamma}\right]\hspace{.8cm}\Pi(g,g;s)=-(p^2g^{\mu\nu}-p^\mu
p^\nu)\left[\frac{4}{9}-\frac{1}{6}\log{p^2\delta
e^\gamma}\right]\eea $\Pi(g,g;g)$ contains the infrared and
collinear divergence (the integral term in the parenthesis), so
the Lorentz covariance is not explicit.

Next we look at the fermion self-mass diagram. If we can still say
that the $D$ dependence in the previous results is explicit so
long as we keep the polarization index $i$ between 1 and $D$, we
now encounter an intrinsic $D$ dependence through the gamma
matrices. So we turn instead to computing the wave function
renormalization $Z_2$. The superficially divergent terms are
calculated to be\bea
\Pi(q,\bar{q})&=&\frac{2}{2-D}\bigg\{-\frac{D}{2}\left[p_{\perp}^i\gamma^i-\frac{p_{\perp}^2}{2p^+}\gamma^+-{p^+}\gamma^-\right]+\frac{(D-8)p^2}{4p^+}\gamma^+\bigg\}\nn\\
&=&\frac{2}{2-D}\left\{\frac{D}{2}\slashed{p}_{\perp}+\frac{(D-8)p^2}{4p^+}\gamma^+\right\}\eea
Here we find it handy to introduce the notation of
$\slashed{p}_{\perp}$, it is an off-shell extrapolation of
$\slashed{p}$ and obeys $\slashed{p}_{\perp}\cdot
\slashed{p}_{\perp}=0$. The result above is not proportional to
$\slashed{p}$, as is necessary to compute $Z_2$. When we try to
embed it into a larger diagram by connecting two propagators to
it\bea
&&\frac{i\slashed{p}}{p^2}\frac{2}{2-D}\left\{\frac{2}{D}\slashed{p}_{\perp}+\frac{(D-8)p^2}{4p^+}\gamma^+\right\}\frac{i\slashed{p}}{p^2}\nn\\
&=&\frac{2}{2-D}\frac{i}{p^2}\left[\slashed{p}_{\perp}+\frac{p^2}{2p^+}\gamma^+\right]\left\{\frac{D}{2}\slashed{p}_{\perp}+\frac{(D-8)p^2}{4p^+}\gamma^+\right\}\left[\slashed{p}_{\perp}+\frac{p^2}{2p^+}\gamma^+\right]\frac{i}{p^2}\nn\\
&=&\frac{2}{2-D}\frac{i}{p^2}\left[\frac{D}{2}\frac{p^4}{2p^+}\gamma^++\frac{(D-8)p^2}{2}\slashed{p}_{\perp}\right]\frac{i}{p^2}%
=\frac{2}{2-D}\frac{i}{p^2}\left[\frac{p^4}{2p^+}\gamma^++\frac{(D-8)p^2}{2}\slashed{p}\right]\frac{i}{p^2}\nn\eea
we must discard the first term as it contains too many powers of
$p^2$ rendering the diagram 1PIR. With this observation we can
simply take $(D-8)/\epsilon$ as the correction to $Z_2$, whose $D$
dependence is differentiable.
\subsection{Triangle Diagrams}
The vertex correction evaluated with a $\delta$-regulator also
shows non-gauge-covariance. For example we observe a mismatch
between $\Gamma(g^i,g^j,+)$ and $\Gamma(g^i,g^j,g^k)$, i.e. a
mismatch when one gluon polarization is changed from $i$ to $-$.
We omit the factor $ig^3N_c/(8\pi^2)\trace{[t^at^bt^c]}$
\bea\Gamma(g^i,g^j,+;s)&=&-(p_1^+-p_2^+)\delta^{ij}\left[\frac{1}{6}\log{p_o^2\delta
e^{\gamma}}-\underline{\frac{5}{18}}\right]\nn\\
\Gamma(g-,g-,g+;s)&=&\frac{-2p_3^+}{p_1^+p_2^+}K^{\wedge}_{2,1}\left[\frac{1}{6}\log{p_o^2\delta
e^{\gamma}}-\underline{\frac{1}{9}}\right]\label{trig_glue_delta}\eea
The two underlined numbers have to be the same to maintain Lorentz
covariance. Next we study how {\dreg} corrects the problem.

The triangle diagrams in the light cone gauge is linearly
divergent, but the same procedure in sec.\ref{tech_dreg} applies
\bea
\Gamma(g^i,g^j,g^k;g)&=&-\left[\frac{2}{2-D}\frac{-2(12+5D)}{3D}+\frac{14}{3}\right]\left\{\delta^{ik}\frac{1}{p_2^+}K^{j}_{21}+\delta^{jk}\frac{1}{p_1^+}K^{i}_{21}+\delta^{ij}\frac{1}{p_3^+}K^{k}_{21}\right\}+\textrm{Art. Div.}\nn\\
\nn\\
\Gamma(g^i,g^j,g^k;q)
&=&-\left[\frac{2}{2-D}\frac{2(2+3D)}{3D}-\frac{10}{3}\right]\left\{\delta^{ik}\frac{1}{p_2^+}K^{j}_{21}+\delta^{jk}\frac{1}{p_1^+}K^{i}_{21}+\delta^{ij}\frac{1}{p_3^+}K^{k}_{21}\right\}\nn\\
\Gamma(g^i,g^j,g^k;s)
&=&-\left[\frac{2}{2-D}\frac{2}{3D}+\frac{1}{3}\right]\left\{\delta^{ik}\frac{1}{p_2^+}K^{j}_{21}+\delta^{jk}\frac{1}{p_1^+}K^{i}_{21}+\delta^{ij}\frac{1}{p_3^+}K^{k}_{21}\right\}\nn\\
%\Gamma(s,s,g^k)&=&\frac{1}{8\pi^2}\left[\frac{2}{2-D}\frac{-4(D+2)}{D}+5\right]\frac{1}{p_3^+}K^{k}_{21}=\left[\frac{2(D+2)}{D}-\frac{5}{2}\right]\Gamma_0
\label{trig_glue_dreg}\eea The first term in each bracket is what
we get by honestly computing in $D$ dimension, the second terms is
due to some peculiarities in using the $\delta$-regulator to
evaluate a linearly divergent diagram. An interested reader may
refer to \cite{Note_by_Thorn} (Eq.108 in particular) for details.
We also give \bea
\Gamma(g^i,g^j,+;s)&=&-(p_1^+-p_2^+)\delta^{ij}\frac{2}{2-D}\left[\frac{D}{12}-\frac{1}{3}\right]\nn\\
\Gamma(g^i,g^j,+;q)&=&-(p_1^+-p_2^+)\delta^{ij}\frac{2}{2-D}\left[-\frac{D}{3}+\frac{4}{3}\right]\eea
$\Gamma(g^i,g^j,g^k;g)$ and $\Gamma(s,s,g^k)$ contain gauge
artificial divergences, which must be cancelled when the self-mass
diagrams on the external legs are included.% Eventually the
%correction 'Dreg'$-\delta$ to the charge renormalization is
%$(D-2)/(6\epsilon)$.

After extracting the finite parts from above, and combining them
with Eq.\ref{trig_glue_delta}, both the $-5/18$ and $-1/9$ become
$-4/9$, restoring gauge covariance.

The fermion vertex correction is more complicated. It can be
separated into one part proportional to $\gamma^{\mu}$, which is
used to compute $Z_1$, and another part of the form
$\gamma^i\gamma^+\slashed{p}$ that vanishes on-shell
%\beaV(q,\bar{q},g^i)=\frac{1}{8\pi^2}\bigg[(4-\frac{D}{2})(-\gamma^+\frac{(p_2-p_4)^i}{(p_2-p_4)^+}+\gamma^i)+\frac{-8+4D-D^2}{4D
%p_1^+}\gamma^i\gamma^+\slashed{p_1}+\frac{8-4D+D^2}{4Dp_2^+}\slashed{p_2}\gamma^+\gamma^i\bigg]\eea
\bea
V(q,\bar{q},g^i)&=&-\left(\frac{2}{2-D}\frac{(D+2)(D-8)}{4D}+\frac{3}{2}\right)\left(\gamma^+\frac{q^i}{q^+}-\gamma^i\right)\nn\\
&&-\left(\frac{2}{2-D}\frac{D^2+2D-16}{8D}+\frac{3}{4}\right)\left(\gamma^i\gamma^+\frac{\slashed{p}_1}{p_1^+}-\frac{\slashed{p}_2}{p_2^+}\gamma^i\gamma^+\right)+\textrm{Art.
Div.}\eea The first term is the tree vertex ($\epsilon_i$ dotted
into $\gamma^{\mu}$), so we will take its coefficient to be the
correction to $1/Z_1$, without a need to interpret the gamma
matrices in non-integer dimension.
%When we include the self-mass diagrams, we
%get the correction to charge renormalization\bea
%\textrm{'Dreg'}-\delta:\hspace{.3cm}
%\frac{\sqrt{Z_3}Z_2}{Z_1}=0\eea free of gauge artificial
%divergence.\footnote{We may do well to remind the reader that this
%has nothing to do with the Ward-Takahashi identity $Z_1=Z_2$, here
%we are computing the difference 'dreg'-$\delta$ of $Z$.}

\subsection{Assembling}\label{Assembling}%
In \cite{II,III}, the helicity conserving gluon amplitudes are
computed with the $\delta$-regulator. We found that the results
have a lot of mismatches such as a lonesome four-point vertex
without the corresponding exchange diagrams to complete to a tree.
In \cite{IV}, when the external states are no longer restricted to
be gluons, the mismatches become more colorful, sometimes an
entire s or t-channel diagram is hanging. When there are only four
particles involved in a (helicity conserving) scattering, the tree
amplitude is the only quantity that satisfies the helicity
requirement and is Lorentz invariant. These can be deduced either
through applying supersymmetry operator to the S-matrix
\cite{GrisaruPendleton} or by brute force exhausting all spinor
expressions one can write down.

We next illustrate schematically how {\dreg} scheme corrects the
mismatch in gluon scattering. The results from \cite{II,III} stand
thus (omitting $ig^4N_c/(8\pi^2)\trace{[t^at^bt^ct^d]}$)\bea
A(\glug,\glug,\gluG,\gluG)_{amputate}&=&\frac{-2K_{12}^{\wedge
4}p_3^+p_4^+}{K_{43}^{\wedge}K_{32}^{\wedge}K_{21}^{\wedge}K_{14}^{\wedge}p_1^+p_2^+}\bigg[-\left(\log^2{\frac{s}{t}}+\pi^2\right)-\frac{11}{3}\log{\delta
e^\gamma{t}}+\frac{73}{9}
+\textrm{IR}\bigg]\nn\\&&-\frac{1}{3}\left(-2\frac{p_1^+p_3^++p_2^+p_4^+}{(p_1^++p_4^+)(p_2^++p_3^+)}\right)+\frac{2}{3}\eea
The first line is proportional to a tree while the second line is
a multiple of four point vertex and a pure number.

As stated in the earlier sections, we should compute all the
superficially divergent parts in this amplitude in 2+D dimension,
and keep the $1/\epsilon$ poles to the very end\bea
\textrm{Superficially divergent}=
-\frac{11}{3}[1/\epsilon+\log{\mu^2}]\textrm{Tree}(2+D)+\textrm{finite}\nn\eea
Now we identify the first term as
$11/3\log{\delta}\cdot\textrm{Tree}(4)$. Those parts that remain
finite as $\epsilon\to 0$ should give $A_{HV}-A_{\delta}$, and we
can evaluate them by taking the dimension back to 4 and go to a
helicity basis\bea
A_{HV}-A_{\delta}=\frac{2}{3}\frac{p_1^+p_3^++p_2^+p_4^+}{(p_1^++p_4^+)^2}-\frac{2}{3}\frac{-2K_{12}^{\wedge
4}p_3^+p_4^+}{K_{43}^{\wedge}K_{32}^{\wedge}K_{21}^{\wedge}K_{14}^{\wedge}p_1^+p_2^+}-\frac{2}{3}\nn\eea
This is to be added to the $\delta$-regulator result, we see that
the hanging four point vertex and the pure number are cancelled.

%Another example is the two gluons scattering into two adjoint
%fermions amplitude. The correction to the contact vertex is\bea
%\frac{i}{6}\frac{p_1^+-p_2^+}{(p_1^++p_2^+)^2}\gamma^+\frac{D-16}{\epsilon}\nn\eea
%The triangle diagrams plus their associated self-mass insertions
%give\bea
%\left[\frac{D-9}{3}\epsilon(p_2)\cdot\gamma\frac{i(p_1+p_4)\cdot\gamma}{(p_1+p_4)^2}\epsilon(p_1)\cdot\gamma+%
%\frac{D-16}{6}(\gamma\cdot\epsilon_i)\frac{iV_{\mu\nu\rho}\epsilon_i^{\mu}\epsilon(p_1)^{\nu}\epsilon(p_2)^{\rho}}{(p_1+p_2)^2}\right]\frac{1}{\epsilon}\nn\eea
%when we go to a helicity basis, we find schematically: \bea
%-\frac{1}{3}\textrm{s-tree}-\frac{2}{3}\textrm{t-tree}\nn\eea
%cancelling the mismatches in \cite{IV}. We shall list all of the
%amplitudes in its entirety after the next section.
%%
\subsection{Dimensional Reduction in the Light
Cone}\label{tech_dred} With the dimensional reduction scheme, all
the tensor algebra or gamma matrix algebra are performed in 4D,
while only the first $4-2\epsilon$ components of all the momenta
are non-zero. To obtain the {\dred} results, we can either insert
$(2\pi)^{2\epsilon}\delta^{2\epsilon}(q)$ into the momentum
integral or build up from the {\dreg} results by including a gluon
whose polarity is in between $D$ and $2$. Its polarization vector
is\bea \epsilon_i^\mu=(0,...,\delta^{\mu}_i,...,0);{\ }i\in [D,2]
\eea actually this gluon behaves in most aspects like an adjoint
scalar (it was called $\epsilon$-scalar in
\cite{CapperJonesNieuwenhuizen}). For example, the four gluon box
with {\dred} scheme up to terms that vanish as $D\to2$ is\bea
B(g^i,g^j,g^k,g^l;g)&=&\frac{2}{2-D}\bigg\{\left[\delta^{il}\delta^{jk}+\delta^{ij}\delta^{kl}\right]\frac{1}{6}+\delta^{ik}\delta^{jl}\frac{10-3D}{6}+\textrm{Art.
Div.}\bigg\}\eea This result can be obtained by combining
$B(g^i,g^j,g^k,g^l;g)$ and $B(g^i,g^j,g^k,g^l;s)(2-D)$ from
Eq.\ref{box_dreg}. $B(g^i,g^j,g^k,g^l;q)$ and
$B(g^i,g^j,g^k,g^l;s)$ remains the same in {\dred}. Here we have
assumed that the external gluon polarization is in $4-2\epsilon$
dimension, for a general external polarization, the results are
given in the appendix. In fact, the calculations there are done
with the tensor algebra or gamma matrix algebra in $D$ dimension
and the momentum integral done in $D_1$ dimension. Then taking
$D=2,D_1=2-2\epsilon$ corresponds to {\dred} and taking
$D=D_1=2-2\epsilon$ corresponds to {\dreg}.

\section {List of Amplitudes}\label{amplitudes}
The scalars and fermions in the following amplitudes are
regrettably given only in the adjoint representation. It turned
out not so trivial to go to a general representation. For example,
for the amplitude of quark pair production
$A(g^a,g^b,q^i,\bar{q}_j)$, the procedure of colour stripping
gives the following structure $(t^at^b)^i_j$, $(t^bt^a)^i_j$ and
$\trace{(t^at^b)}\delta^i_j$. So it is necessary to include the
diagrams where the two gluons legs are crossed, and thereby lose
planarity. Looking back to our calculation done in
\cite{I,II,III,IV}, the detailed cancellation of the gauge
artificial divergence depended rather delicately on having a
planar representation of the Feynman diagrams. It is only after
the cancellation of the artificial divergence do we find the
amplitude together with its infrared terms symmetric under
crossing.

The one loop amplitudes are all proportional to the tree level\bea
A_1=\frac{-g^2N_c}{16\pi^2}A_0\times F\eea and in each expression
of $F$ in the list, the upper entry in the array applies to
{\dreg} and the lower to {\dred} scheme.
\bea F(\glug,\glug,\gluG,\gluG)&=&
\left[\begin{array}{c}67/9\\64/9\end{array}\right]-\frac{4}{9}N_s-\frac{20}{9}N_f\nn\\
&&\left[-\frac{11}{3}+\frac{1}{6}N_s+\frac{4}{3}N_f\right]\left[-\frac{1}{\epsilon}+\log{\frac{te^\gamma}{\mu^2}}\right]-\left(\log^2{\frac{s}{t}}+\pi^2\right)\nn\\
A(\glug,\glug,\gluG,\gluG)_0&=&\pref\frac{-2K_{12}^{\wedge
4}p_3^+p_4^+}{K_{43}^{\wedge}K_{32}^{\wedge}K_{21}^{\wedge}K_{14}^{\wedge}p_1^+p_2^+}\to\pref\frac{-2\bra{1}2\rangle^4}{\bra{1}2\rangle
\bra{2}3\rangle \bra{3}4\rangle \bra{4}1\rangle}\nn\\
F(\glug,\gluG,\glug,\gluG)&=&\left[\begin{array}{c}67/9\\64/9\end{array}\right]-\frac{4}{9}N_s-\frac{20}{9}N_f+\frac{st(2+N_s-4N_f)}{2(s+t)^2}\nn\\
&&+\left[-\frac{11}{3}+\frac{1}{6}N_s+\frac{4}{3}N_f\right]\left[-\frac{1}{\epsilon}+\log{\frac{se^\gamma}{\mu^2}}\right]+\frac{s}{6(s+t)^3}\log{\frac{s}{t}}\times\nn\\&&\left[(28t^2+38st+22s^2)+N_s(2t^2-5st-s^2)-N_f(8s^2+4st+20t^2)\right]\nn\\
&&-\frac{1}{2(t+s)^4}\left[2(s^2+st+t^2)^2+2N_f(st^3+s^3t)+N_s
s^2t^2\right]\left(\log^2{\frac{s}{t}}+\pi^2\right)\nn\\
A(\glug,\gluG,\glug,\gluG)_0&=&\pref\frac{-2K_{13}^{\wedge
4}p_2^+p_4^+}{K_{43}^{\wedge}K_{32}^{\wedge}K_{21}^{\wedge}K_{14}^{\wedge}p_1^+p_3^+}\to\pref\frac{-2\bra{1}3\rangle^4}{\bra{1}2\rangle
\bra{2}3\rangle \bra{3}4\rangle \bra{4}1\rangle}\nn\\\eea

\bea
F(s,s,\glug,\gluG)&=&\left[\begin{array}{c}8\\8\end{array}\right]-4\left[-\frac{1}{\epsilon}+\log{\frac{se^\gamma}{\mu^2}}\right]+\frac{2s}{s+t}\log{\frac{s}{t}}
-\frac{s^2+st+t^2}{(s+t)^2}\left(\log^2{\frac{s}{t}}+\pi^2\right)\nn\\
A(s,s,\glug,\gluG)_0&=&\pref\frac{-8K_{13}^{\wedge 2}K_{14}^{\vee
2}}{st
p_3^+p_4^+p_1^{+2}}\to\pref\frac{-2\bra{3}1\rangle^2[1|4]^2}{st}\nn\\
F(\glug,s,\gluG,s)&=&\left[\begin{array}{c}8\\8\end{array}\right]-4\left[-\frac{1}{\epsilon}+\log{\frac{se^\gamma}{\mu^2}}\right]+2\log{\frac{s}{t}}-\left(\log^2{\frac{s}{t}}+\pi^2\right)\nn\\
A(\glug,s,\gluG,s)_0&=&\pref\frac{-8K_{21}^{\wedge 2}K_{32}^{\vee
2}}{st
p_1^+p_3^+p_2^{+2}}\to\pref\frac{-2\bra{1}2\rangle^2[3|2]^2}{st
}\nn\\
F(s,s,\fl{},\Fl{})&=&\left[\begin{array}{c}68/9\\62/9\end{array}\right]+\frac{4}{9}N_s+\frac{20}{9}N_f\nn\\
&&-\left[\frac{10}{3}+\frac{1}{6}N_s+\frac{4}{3}N_f\right]\left[-\frac{1}{\epsilon}+\log{\frac{s
e^{\gamma}}{\mu^2}}
\right]-\frac{2t+s}{2(s+t)}\left(\log^2{\frac{s}{t}}+\pi^2\right)\nn\\
A(s,s,\fl{},\Fl{})_0&=&\pref\frac{4K_{13}^{\wedge}K_{14}^{\vee}}{s
p_1^+p_3^+}\to\pref\frac{2\bra{3}1\rangle[1|4]}{s}\eea%

\bea A(s_\alpha,s_\beta,s_\beta,s_\alpha)_1&=&\frac{ig^4N_c}{8\pi^2}\trace(t^at^bt^ct^d)\bigg[\frac{1}{9}\left[\begin{array}{c}95+154s/t\\107+160s/t\end{array}\right]\nn\\&&+\left(\frac{2s}{t}+1\right)\left[\frac{4}{9}N_s+\frac{20}{9}N_f-\left(\frac{1}{6}N_s+\frac{4}{3}N_f\right)\left[-\frac{1}{\epsilon}+\log{\frac{te^{\gamma}}{\mu^2}}\right]\right]\nn\\
&&-\frac{2(13s+8t)}{3t}\left[-\frac{1}{\epsilon}+\log{\frac{te^{\gamma}}{\mu^2}}\right]-\frac{1}{2}\left[-\frac{1}{\epsilon}+\log{\frac{se^{\gamma}}{\mu^2}}\right]-\frac{2s}{t}\left(\log^2{\frac{s}{t}}+\pi^2\right)\bigg]\nn\\
A(s_\alpha,s_\beta,s_\beta,s_\alpha)_0&=&\pref\frac{2s+t}{t};\;(\alpha\neq
\beta)\eea

\bea
F(\glug,\gluG,\fl{},\Fl{})&=&\left[\begin{array}{c}7\\6\end{array}\right]-3\left[-\frac{1}{\epsilon}+\log{
\frac{se^{\gamma}}{\mu^2}
}\right]+\frac{3s}{s+t}\log{\frac{s}{t}}-\frac{2s^2+st+2t^2}{2(s+t)^2}\left(\log^2{\frac{s}{t}}+\pi^2\right)\nn\\
A(\glug,\gluG,\fl{},\Fl{})_0&=&\pref\frac{-8K_{13}^{\wedge
2}K_{32}^{\vee}K_{24}^{\vee}}{st
p_1^+p_2^+p_3^{+2}}\to\pref\frac{-2\bra{3}1\rangle^2
[3|2][2|4]}{st}\nn\\
F(\glug,\gluG,\fr{},\Fr{})&=&\left[\begin{array}{c}7\\6\end{array}\right]-3\left[-\frac{1}{\epsilon}+\log{
\frac{se^{\gamma}}{\mu^2}
}\right]-\left(\log^2{\frac{s}{t}}+\pi^2\right)\nn\\
A(\glug,\gluG,\fr{},\Fr{})_0&=&\pref\frac{8K_{14}^{\wedge
}K_{13}^{\wedge}K_{32}^{\vee 2}}{st
p_1^+p_2^+p_3^{+2}}\to\pref\frac{2\bra{4}1\rangle\bra{3}1\rangle
[3|2]^2}{st}\nn\\
A(\glug,\fl{},\gluG,\Fl{})&=&\left[\begin{array}{c}7\\6\end{array}\right]-3\left[-\frac{1}{\epsilon}+\log{
\frac{te^{\gamma}}{\mu^2}
}\right]-\left(\log^2{\frac{s}{t}}+\pi^2\right)\nn\\
A(\glug,\fl{},\gluG,\Fl{})_0&=&\pref\frac{-8K_{21}^{\wedge
2}K_{43}^{\vee}K_{32}^{\vee}}{st
p_1^+p_3^+p_2^{+2}}\to\pref\frac{-2\bra{1}2\rangle^2[4|3][3|2]}{st}
\eea%

\bea
F(\fl{\alpha},\fl{\beta},\Fl{\beta},\Fl{\alpha})&=&\left[\begin{array}{c}59/9\\44/9\end{array}\right]+\frac{4}{9}N_s+\frac{20}{9}N_f
-\left[\frac{7}{3}+\frac{1}{6}N_s+\frac{4}{3}N_f\right]\left[-\frac{1}{\epsilon}+\log{
\frac{te^{\gamma}}{\mu^2}
}\right]\nn\\&&-\left(\log^2{\frac{s}{t}}+\pi^2\right)\nn\\
A(\fl{\alpha},\fl{\beta},\Fl{\beta},\Fl{\alpha})_0&=&\pref\frac{-4K_{21}^{\wedge
}K_{43}^{\vee}}{t
p_1^+p_2^+}\to\pref\frac{-2\bra{1}2\rangle[4|3]}{t};\;(\alpha\neq \beta)\nn\\
F(\fl{\alpha},\Fl{\beta},\fl{\beta},\Fl{\alpha})&=&\left[\begin{array}{c}59/9\\44/9\end{array}\right]+\frac{4}{9}N_s+\frac{20}{9}N_f
-\left[\frac{7}{3}+\frac{1}{6}N_s+\frac{4}{3}N_f\right]\left[-\frac{1}{\epsilon}
+\log{\frac{te^{\gamma}}{\mu^2}}\right]\nn\\&&-\frac{t}{s+t}\log{\frac{s}{t}}
-\frac{2s^2+st+2t^2}{2(s+t)^2}\left(\log^2{\frac{s}{t}}+\pi^2\right)\nn\\
A(\fl{\alpha},\Fl{\beta},\fl{\beta},\Fl{\alpha})_0&=&\pref\frac{4K_{13}^{\wedge
}K_{24}^{\vee}}{t
p_1^+p_3^+}\to\pref\frac{2\bra{3}1\rangle[2|4]}{t};\;(\alpha\neq \beta)\nn\\
F(\fl{\alpha},\Fl{\alpha},\fl{\beta},\Fl{\beta})&=&\left[\begin{array}{c}59/9\\44/9\end{array}\right]+\frac{4}{9}N_s+\frac{20}{9}N_f
-\left[\frac{7}{3}+\frac{1}{6}N_s+\frac{4}{3}N_f\right]\left[-\frac{1}{\epsilon}+\log{
\frac{se^{\gamma}}{\mu^2}}\right]\nn\\&&+\frac{s}{s+t}\log{\frac{s}{t}}
-\frac{2s^2+st+2t^2}{2(s+t)^2}\left(\log^2{\frac{s}{t}}+\pi^2\right)\nn\\
A(\fl{\alpha},\Fl{\alpha},\fl{\beta},\Fl{\beta})_0&=&\pref\frac{4K_{13}^{\wedge
}K_{24}^{\vee}}{s
p_1^+p_3^+}\to\pref\frac{2\bra{3}1\rangle[2|4]}{s};\;(\alpha\neq
\beta)\eea In giving the tree amplitude, we have converted the
$K_{ij}$ into the {\it normalized} spinors to conform to the
fashion. The reader may find that in an amplitude involving
fermions the factors of $p^+$ is not what is needed to convert the
$K_{ij}$ into the spinor products. The reason is merely for the
sake of automated calculation. The correct external line factor is
$\sqrt{\sqrt{2}p_i^+}$ for every fermion line, but in order to
avoid the phase ambiguity, we only associate $\sqrt{2}p_i^+$ to an
outgoing fermi line.

\section {N=1 Supersymmetry}\label{supersymmetry}
One can go to $N=1$ gauge multiplet by taking $N_s=0$ and
$N_f=1/2$. The amputated amplitudes with {\dred} are given by\bea
A(-,-,+,+)&=&tree_1
\bigg[6-3\left[-\frac{1}{\epsilon}+\log{\frac{te^\gamma}{\mu^2}}\right]-\left(\log^2{\frac{s}{t}}+\pi^2\right)\bigg]\nn\\
A(-,+,-,+)&=&tree_2\bigg[6-3\left[-\frac{1}{\epsilon}+\log{\frac{se^\gamma}{\mu^2}}\right]+\frac{3s}{(s+t)}\log{\frac{s}{t}}-\frac{(2s^2+st+2t^2)}{2(t+s)^2}\left(\log^2{\frac{s}{t}}+\pi^2\right)\bigg]\eea
From Eq.\ref{ALL_sfmass}, the self-mass is already supersymmetric,
we need only check the supersymmetry relations among the tree
amplitudes here. The covariant treatment can be found in
\cite{ZAZ}. Here we stick to the light cone and write down the
trees with the light cone superfields. Indeed, the two amplitudes
above correspond to $A(\phi,\phi,\bar{\phi},\bar{\phi})$ and
$A(\phi,\bar{\phi},\phi,\bar{\phi})$ respectively. Referring to
the appendix, the four point vertex can be absorbed into the
exchange diagrams, so only three graphs contribute to the first
amplitude (a factor of $-2g^2\trace{[t^at^bt^ct^d]}$ is understood
to multiply the following)
\begin{figure}[h]
\begin{center}
\psfrag {a,p1}{$a,\,p_1$}\psfrag {b,p2}{$b,\,p_2$}\psfrag
{c,p3}{$c,\,p_3$}\psfrag {d,p4}{$d,\,p_4$}
\includegraphics[width=2.5in]{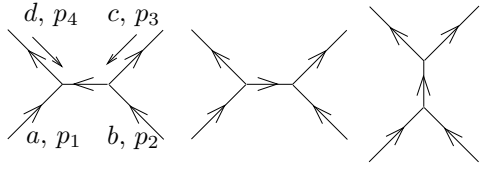}
\caption{The three Feynman diagrams that contribute to
$A(\phi,\phi,\bar{\phi},\bar{\phi})$, incoming (outgoing) arrow
corresponds to $\phi$ ($\bar{\phi}$)}\label{sf_tree_eps}
\end{center}
\end{figure}

Fig.\ref{sf_tree_eps} gives\bea
A(\phi,\phi,\bar{\phi},\bar{\phi})&=&i\int{d\theta}\bigg\{(D\bar{\phi}_4)\phi_1D(\bar{\phi}_3D\phi_2)\langle
1|4\rangle[4|3]\frac{p_4^+}{t(p_1+p_4)^+}\nn\\
&+&(D\bar{\phi}_3)\phi_2D(\bar{\phi}_4D\phi_1)\langle
2|3\rangle[3|4]\frac{p_3^+}{t(p_2+p_3)^+}-D(\bar{\phi}_4\bar{\phi}_3)\phi_1\phi_2\langle
1|2\rangle[3|4]\frac{2(p_1^++p_2^+)}{s}\bigg\}\nn\eea%
where $\phi_i$ means $\phi(p_i;\theta)$ and the spinors are
unnormalized spinors. We can further simplify the above to get\bea
iD\bigg\{ D(\bar{\phi}_4\bar{\phi}_3)\phi_1\phi_2\frac{\langle
12\rangle[34]}{st(p_1+p_2)^+}\bigg[-4p_1^+p_2^+p_3^+p_4^+\langle
12\rangle[34]-2sp_1^+p_3^+\bigg]
-(D\bar{\phi}_3)\phi_2D(\bar{\phi}_4D\phi_1)\frac{\langle
12\rangle[34]}{t}\bigg\}\bigg|\eea By inserting the above in
between definite incoming and outgoing asymptotic states, one can
check that it does give the correct component amplitudes. In fact
the verification is trivial for the case of four gluon and four
gluino scattering. The verification of two gluon and two gluino
amplitude is a bit involved. As most readers are not familiar with
manipulating light cone spinors, we shall demonstrate it as an
example. If we aim for $A(\fl{},\glug,\gluG,\Fl{})$, then we
arrange all the $D$'s to act on $\phi_1$ and $\bar{\phi}_4$\bea
A(\fl{},\glug,\gluG,\Fl{})&=&\bigg\{
-(D\bar{\phi}_4)\bar{\phi}_3(D\phi_1)\phi_2\frac{i\langle
12\rangle[34]}{st(p_1+p_2)^+}\bigg[-4p_1^+p_2^+p_3^+p_4^+\langle
12\rangle[34]-2sp_1^+p_3^+\bigg]\nn\\
&-&(2p_3^+\bar{\phi}_3)\phi_2(D\bar{\phi}_4)(D\phi_1)\frac{i\langle
12\rangle[34]}{t}\bigg\}\bigg|\nn\\
&=&-\sqrt{2}\bar{\psi}_4A^{\wedge}_3\psi_1A^{\vee}_2\bigg\{-\frac{i\langle
12\rangle^2[34]^2}{st(p_1+p_2)^+}4p_1^+p_2^+p_3^+p_4^+{}
+2p_2^+p_3^+\frac{i\langle
12\rangle[34]}{t(p_1+p_2)^+}\bigg\}\label{before_swap}\eea Use the
identity\bea p_1^+p_2^+p_4^+\langle
2|1\rangle=(p_1+p_2)^+p_2^+p_4^+\langle
2|4\rangle-p_2^+p_3^+p_4^+\langle 4|3\rangle\nn\eea to swap one of
the $\langle 12\rangle$ in the first term of
Eq.\ref{before_swap}\bea\frac{i\langle
12\rangle^2[34]^2}{st(p_1+p_2)^+}4p_1^+p_2^+p_3^+p_4^+&=&\frac{i\langle
12\rangle[34]^2}{st(p_1+p_2)^+}4p_3^+\left[-(p_1+p_2)^+p_2^+p_4^+\langle
2|4\rangle+p_2^+p_3^+p_4^+\langle 4|3\rangle\right]\nn\\
&=&\frac{-4i\langle 12\rangle\langle
24\rangle[34]^2}{st}p_3^+p_2^+p_4^++\frac{2i\langle
12\rangle[34]}{st(p_1+p_2)^+}p_3^+p_2^+\eea%
we arrive at\bea
A(\fl{},\glug,\gluG,\Fl{})&=&-\sqrt{2}\bar{\psi}_4A^{\wedge}_3\psi_1A^{\vee}_2\frac{4i\langle
12\rangle\langle 24\rangle[34]^2p_3^+p_2^+p_4^+}{st}\nn\eea By
inserting the above into incoming and outgoing states, we obtain
an extra factor of $\sqrt{2p_1^+p_4^+}$ from the free field
expansion of $\psi$. So\bea
A(\fl{},\glug,\gluG,\Fl{})=-\frac{8i\langle 12\rangle\langle
24\rangle[34]^2p_3^+p_2^+p_4^+\sqrt{p_1^+p_4^+}}{st}\rightarrow-\frac{2i\langle
12\rangle\langle 24\rangle[34]^2}{st}\eea The spinors in the last
expression are properly normalized.

Similarly, the amplitude $A(\phi,\bar{\phi},\phi,\bar{\phi})$ can
be written\bea
A(\phi,\bar{\phi},\phi,\bar{\phi})&=&iD\bigg\{\left(-(D\bar{\phi}_2)\phi_1D(\bar{\phi}_4D\phi_3)\frac{1}{s}+(D\bar{\phi}_2)\phi_3D(\bar{\phi}_4D\phi_1)\frac{1}{t}\right)\langle
1|3\rangle[2|4]\nn\\
&-&D(\bar{\phi}_2\bar{\phi}_4)\phi_1\phi_3\frac{\langle
1|3\rangle[2|4]}{st(p_1+p_3)^+}\left(4p_1^+p_2^+p_3^+p_4^+\langle
1|3\rangle[2|4]-2sp_1^+p_2^++2tp_2^+p_3^+\right)\bigg\}\bigg|\nn\eea

\section{Infrared Terms}\label{Infrared Terms}%
So far we have been silent about all the infrared terms and
external self-mass diagrams. The rigorous treatment of the
infrared issue can be found in \cite{II}. Here we merely try to
establish the link between the infrared terms regulated with cut
off in $q^+$ and the ones regulated by an almighty $\epsilon$ in
the literature.

One can always factor a scattering amplitude into the following
form \cite{IR_factor,StermanTejedaYeomans}\bea
A_n=J\left(\frac{Q^2}{\mu^2},\alpha(\mu)\right)\times
S\left(p_i,\frac{Q^2}{\mu^2},\alpha(\mu)\right)\times
h_n\left(p_i,\frac{Q^2}{\mu^2},\alpha(\mu)\right)\label{factorization}\eea
where $J$ is the 'jet' function and $S$ is the 'soft' part and
$h_n$ is the hard remainder, which is finite after renormalization
is performed. In a dimensional regulation scheme, the jet function
will contain a double pole in $\epsilon$, caused by the overlap of
collinear divergence and infrared divergence. In the light cone
gauge, it is given by the external self-mass terms. In fact, the
collinear divergence shows up as $\log{p^2}$ in
Eq.\ref{ALL_sfmass}, and the infrared divergence shows up as the
divergent integral in $q^+$. $\log{p^2}$ can be replaced with
$\log{(\Delta^2)}$ ($\Delta$ is the jet resolution) when the
collinear emissions are included.

The rest of the infrared terms are interpreted as the soft part
$S$ above. They can be cast into the following form\bea \textrm{IR
term}&=&-\frac{ig^2N_c}{16\pi^2}A_0\sum_{i=1...4} I(p_i,p_{i+1})\nn\\
I(p_i,p_{i+1})&=&\int_{0}^{|p_{i}^+|}{\frac{dq^+}{q^+}\left[-\frac{1}{\epsilon}+\log{\frac{q^{+2}|(p_i+p_{i+1})^2|e^\gamma}{\mu^2|p_i^+p_{i+1}^+|}}\right]}+\int_{0}^{|p_{i+1}^+|}{\frac{dq^+}{q^+}\left[-\frac{1}{\epsilon}+\log{\frac{q^{+2}|(p_i+p_{i+1})^2|e^\gamma}{\mu^2|p_i^+p_{i+1}^+|}}\right]}\nn\\
&+&\int_{-p_i^+}^{p_{i+1}^+}{\frac{dq^+}{q^+}\log{\left|\frac{(p_{i+1}^+-q^+)p_i^+}{(q^++p_{i}^+)p_{i+1}^+}\right|}}\label{IR}\eea
First of all, the appearance of $\epsilon$ in Eq.\ref{IR} is
temporary, our IR singularity is {\textit {not}} regulated by
$\epsilon$. The form of the infrared terms are slightly
asymmetric: the third integral might or might not cross over 0
depending on the sign of the momenta, giving a difference
proportional to $\pi^2$. We have not computed the amplitudes with
quarks in the fundamental representation, but we believe that in
general there is going to be one such $I(p_i,p_{i+1})$ for the
region between particle i and i+1 if it is bounded by a color
line. So
\begin{figure}[h]
\begin{center}
\includegraphics[width=1.2in]{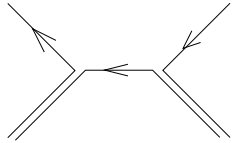}
\caption{The double line notation for the quark pair production
amplitude}\label{sudakov}
\end{center}
\end{figure} for the amplitude fig.\ref{sudakov}, there will be
$I(p_1,p_{2})$, $I(p_2,p_{3})$ and $I(p_4,p_{1})$ but no
$I(p_3,p_{4})$. The reason is that these IR terms are nothing but
Sudakov form factors, which is the square root of the photon to
quark anti-quark pair amplitude.

We need to establish the dictionary between the light cone
infrared terms and the dimensionally regulated infrared term. This
is simply achieved by subtracting the amplitudes given in
sec.\ref{amplitudes} by the corresponding ones obtained with the
coventional dimensional regulation. The non-trivial point is that
the dictionary thus obtained should be universal. The amplitude
for MSYM is the simplest, it is given by\bea
A(g,g,g,g;\textrm{\small
MSYM})=\frac{ig^2N_c}{16\pi^2}\left(\log^2{\frac{s}{t}}+\pi^2\right)A_{tree}+\textrm{ext
self-mass+IR terms}\eea the same is given in \cite{BDS} with
dimensional reduction\bea A(g,g,g,g;\textrm{\small
MSYM})&=&-\frac{1}{2}stI_4^{(1)}(s,t)\frac{ig^2N_c}{8\pi^2}A_{tree}\nn\\
&=&\frac{ig^2N_c}{8\pi^2}\left[-\frac{2}{\epsilon^2}+\frac{1}{\epsilon}\log{\frac{st}{\mu^4}}-\log{\frac{s}{\mu^2}}\log{\frac{t}{\mu^2}}+\frac{2\pi^2}{3}\right]A_{tree}\eea
The difference of the two expressions above gives the
correspondence between the infrared terms\bea
\textrm{Eq.\ref{IR}}&=&-\frac{\pi^2}{3}+\frac{4}{\epsilon^2}+\frac{2}{\epsilon}\log{\frac{\mu^4}{st}}+\log^2{\frac{\mu^2}{s}}+\log^2{\frac{\mu^2}{t}}\nn\\&&+\sum_i\int_0^{p_i^+}{\frac{dq^+}{p_i^+}\left[\frac{-2}{p_i^+-q^+}+\frac{-2}{q^+}\right]\left[-\frac{1}{\epsilon}+\log{\frac{(p_i^+-q^+)q^+\Delta^2
e^\gamma}{\mu^2p_i^{+2}}}\right]}\label{dic_IR}\eea Note we can
reshuffle the $1/\epsilon$ terms from the last term (external
self-mass terms) on the rhs to lhs. In fact, they precisely
cancel, as promised underneath Eq.\ref{IR}. We can use the
dictionary Eq.\ref{dic_IR} to compare the rest of our results in
sec.\ref{amplitudes} with those in the literature.

Eq.\ref{dic_IR} is also interesting in the sense that it can be
checked by computing a light-like Wilson loop with cusps, but with
the infrared divergence regulated with our scheme.

\section{Concluding Remarks}
To summarize, in this article we focused on obtaining the
difference in the amplitudes between {\dreg}, {\dred} and the
$\delta$-regulator. The computation is done in a cartesian basis,
contrary to what is in fashion. The dimensional parameter
$\epsilon$ is only used to regulate the UV, while the IR is
regulated by cutting off $q^+$, as was done in our earlier works.
In the context of $N=4$ SYM, this means the total disappearance of
$\epsilon$. In this aspect, our regularization scheme is quite
different from the conventional dimensional regularization
schemes, put in less accurate words, our calculation is as close
to 4D as it can get and yet still preserves gauge symmetry. The
supersymmetric identities in $N=1$ SUSY between various amplitudes
can be easily checked. The results using {\dreg} of course shows
SUSY violation.

The IR terms thus obtained is universal as they should be, in
fact, we see from Eq.\ref{dic_IR} that can be interpreted as the
Sudakov form factor. As a subject for further investigation, we
would like to understand the correspondence Eq.\ref{dic_IR}
better. On one hand our IR terms are well 'localized', that is,
involves two neighboring legs at a time, on the other hand, we
know in conventional dimensional regulation the external self-mass
terms vanish and the IR singularity comes from a single scalar box
integral\bea
I_4^{(1)}(s,t)=-ie^{\epsilon\gamma}\pi^{-(2-\epsilon)}\int{\frac{d^{4-2\epsilon}q}{q^2(q-p_1)^2(q-p_1-p_2)^2(q+p_4)^2}}\eea
We have not yet studied how to obtain the correspondence
Eq.\ref{dic_IR} from a presumably short calculation. In
\cite{StermanTejedaYeomans,MagneaSterman}, it was shown that this
form factor exponentiates, which is a prototype of the BDS
iteration relation. With the current data in the light cone at
hand, we are far from being able to study the exponentiation of
the IR terms. One possible way to proceed is to clarify the
correspondence Eq.\ref{dic_IR} further and convert the IR term
from the literature into light cone form. As was mentioned in the
introduction, we can shuffle finite terms among the three factors
in Eq.\ref{factorization}. With our simple minded infrared
regulation, the hard remainder $h_n$ (listed in
sec.\ref{amplitudes}) is fortunately Lorentz covariant. We take it
as a good sign that the factorization ordained by our infrared
regulation scheme will give a hard remainder that also obeys a
nice iteration relation similar to \cite{BDS}. It is presumably
quite hard to push the light cone computation to high loops, no
matter how much we advertise it, it is not the best scheme to do
calculations with. Alternative routes are yet to be found.

\noindent {\bf Acknowledgements} The author would like to thank
prof C.B.Thorn for both initiating the current work and giving a
lot of helpful consultations along the way. The work is supported
by the grant for research of the Department of Theoretical
Physics, Uppsala University.
\appendix
\section{Notation and Convention}\textit{Metric} in this article is $g^{\mu\nu}=diag(1,-1,-1,-1)$.

\noindent \textit{Polarization vectors} are given by\bea
\epsilon_i^\mu=(\frac{1}{\sqrt{2}}\frac{k^i}{k^+},...,\delta^{\mu}_i,...,-\frac{1}{\sqrt{2}}\frac{k^i}{k^+})\eea
satisfies\bea \sum_i{\epsilon_i^{\mu}\epsilon_i^{\nu}}+g^{\mu
+}g^{\nu +}\frac{k^2}{k^{+2}}=-g^{\mu\nu}+\frac{g^{\mu
+}k^{\nu}}{k^+}+\frac{g^{\nu +}k^{\mu}}{k^+}\eea The rhs is the
numerator of the gluon propagator, so it can be decomposed into a
part that describes the transverse polarizations and a part that
describes the 'propagation of the $A^-$. The $k^2$ factor on the
lhs will cancel the propagator, which shows that $A^-$ is
non-dynamical. The $1/k^+$ factor is what we call 'gauge
artificial divergence' throughout the article. The fate of these
divergences in a physical quantity is:\begin{itemize}\item simple
and double poles must all cancel\item double poles coincident with
a logarithm must cancel\item simple poles coincident with a
logarithm must arrange themselves into the universal IR terms
Eq.\ref{IR}\end{itemize}

Using the polarization vectors we can pass from the covariant YM
lagrangian to the light cone gauge fixed lagrangian. A three point
vertex in the light cone can be obtained by dotting these
polarization vectors into the covariant vertex functions. The
results are simply\bea
%%V^{ijk}=\delta^{ij}\left[(p_1-p_2)^+\frac{p_3^k}{p_3^+}-(p_1-p_2)^k\right]%
%%+\delta^{jk}\left[(p_2-p_3)^+\frac{p_1^i}{p_1^+}-(p_2-p_3)^i\right]%
%%+\delta^{jk}\left[(p_3-p_1)^+\frac{p_2^j}{p_2^+}-(p_3-p_1)^j\right]\eea
V^{ijk}=\delta^{ij}\left[(p_1-p_2)^+\frac{p_3^k}{p_3^+}-(p_1-p_2)^k\right]+\textrm{cyclic
permutation}\eea By fixing, say, $i=j=\wedge$ and $k=\vee$, we may
reduce the three terms above to only one. This shows the advantage
of the helicity basis over the cartesian basis.

\noindent \textit{Spinors} in the light cone are given by dotting
a $\slashed{p}$ into a reference spinor $\eta$. \bea
&&\ket{p}=-\frac{1}{\sqrt{2}p^+}(p_{\mu}\sigma^{\mu})^{a\dot{a}}\eta_{\dot{a}}=-\frac{1}{\sqrt{2}p^+}p_{\mu}\sigma^{\mu}\left[\begin{array}{c}0\\-1\end{array}\right]=\left[\begin{array}{c}-\frac{p^{\wedge}}{p^+}\\1\end{array}\right];\
[p|=\ket{p}^{\dag}\nn\\
&&|p]=\frac{1}{\sqrt{2}p^+}(p_{\mu}\bar{\sigma}^{\mu})_{\dot{a}a}\eta^a=\frac{1}{\sqrt{2}p^+}p_{\mu}\bar{\sigma}^{\mu}\left[\begin{array}{c}1\\0\end{array}\right]=\left[\begin{array}{c}1\\
\frac{p^{\vee}}{p^+}\end{array}\right];\ \bra{p}=|p]^{\dag}\eea
These spinors are not properly normalized unless we further
multiply them by $2^{1/4}\sqrt{p^+}$. The spinors appearing in
this article are all unnormalized unless stated otherwise.

The quantity $K_{i,j}$ was also used in our previous calculations.
They are directly related to the spinor products and we give them
here to facilitate the comparison with our earlier works\bea
K^\wedge_{q,p}&=&\bra{p}q\rangle{p^+q^+}=\left(q^+p^\wedge
-p^+q^\wedge\right)\nn\\
K^\vee_{p,q}&=&[p|q]{p^+q^+}=\left(p^+q^\vee -q^+p^\vee\right)\eea
we use the unnormalized spinors or the $K$'s, to avoid the phase
ambiguity in $\sqrt{p^+}$.

\noindent \textit{Color Stripping} decomposes an amplitude into
color subamplitudes\bea
M(1,2...,n)=\Sigma_{\sigma}{\trace{[t^{\sigma(1)}t^{\sigma(2)}...t^{\sigma(n)}]}}A(\sigma(1),\sigma(2),...\sigma(n))\eea
where $t^a$ is in the fundamental representation normalized so
that $\trace{t^at^b}=1/2\delta^{ab}$, and summation is over all
the cyclically inequivalent permutation $\sigma$. Since we have
restricted ourselves to the fields in the adjoint representation,
this procedure is equivalent to picking out all the planar
diagrams with the external legs ordered in the specific order
given by $\sigma$.

Lastly, the usual factors associated with dimensional regulation
such as $(2\pi)^{-\epsilon}$ from the momentum integration and
$2^{-\epsilon}$ from $\trace{1}$ will be left out to avoid
cluttering the presentation. These factors, if necessary, can be
recovered easily.

\section{Light cone superfield} We follow Mandelstam
\cite{Mandelstam} and use only one real Grassmann coordinate for
one supersymmetry. Define the super derivative\bea
D=\frac{\partial}{\partial\theta}+2i\partial^+\theta;\ D^2=2i\partial^+\eea%
The superfields are given by\bea \phi=A^{\vee}+2^{1/4}\theta \psi
;\ \bar{\phi}=A^{\wedge}+2^{1/4}\theta \bar{\psi}\eea $\bar{\phi}$
is \textit{not} the conjugate of $\phi$. The N=1 gauge multiplet
can be packed nicely by using spinor notations\bea
L&=&[\bar{\phi}^a|\left\{i\frac{1}{\sqrt{2}}\delta^{ac}\partial\cdot\bar{\sigma}-igf^{abc}\left(\frac{1}{2}|\bar{\phi}^b]\bra{\eta}+|\eta]\bra{\phi^b}\right)\right\}|D\phi^c\rangle\nn\\
&-&g^2f^{abe}f^{ecd}[D\phi^a|\eta]\bra{\phi^b}\eta\rangle\frac{D}{-2\partial^{+2}}[\phi^c|\eta]\bra{D\phi^d}\eta\rangle\eea
where \bea
\ket{\phi}=\left[\begin{array}{c}-\frac{\partial^\wedge}{\partial^+}\\1\end{array}\right]\phi;\hspace{.5cm}%
\bra{\phi}=\left[\begin{array}{c}1\\\frac{\partial^\wedge}{\partial^+}\end{array}\right]\phi;\hspace{.5cm}%
|\bar{\phi}]=\left[\begin{array}{c}1\\ \frac{\partial^\vee}{\partial^+}\end{array}\right]\bar{\phi};\hspace{.5cm}%
[\bar{\phi}|=\left[\begin{array}{c}-\frac{\partial^\vee}{\partial^+}\\1\end{array}\right]\bar{\phi}\eea%
The propagator can be read
off\bea\phi^a\bar{\phi}^b\sim\delta^{ab}\frac{2\partial^+}{\partial^2
D}\eea%
or equivalently
\bea\ket{\phi^a}[\bar{\phi}^b|\sim\delta^{ab}\frac{2}{\partial^2
D}\left[\begin{array}{cc}\frac{\partial^\vee\partial^\wedge}{\partial^+}&-\partial^\wedge\\-\partial^{\vee}&\partial^+\end{array}\right]=\delta^{ab}\frac{2}{\partial^2
D}\left(\frac{1}{\sqrt{2}}\slashed{\partial}-\frac{\partial^2}{2\partial^+}\ket{\eta}[\eta|\right)\eea%
The quartic term is written to look like two three-point vertices
connected by a propagator. We observe that the effect of it is
precisely to cancel the second term in the propagator. So we
ignore the quartic term and
use\bea\ket{\phi^a}[\bar{\phi}^b|\sim\delta^{ab}
\frac{\sqrt{2}}{\slashed{\partial}D}\eea as the propagator from
now on. This observation is valid not only for the four point
amplitudes, but for any Feynman diagrams.
\section{Some Intermediate Results}
\noindent \textbf{Self Mass Diagrams}

The upper entry applies to {\dreg} and the lower to {\dred}
schemes\bea\Pi(s,s)&=&p^2\bigg\{\left[\begin{array}{c}-8\\-8\end{array}\right]+4\left[-\frac{1}{\epsilon}+\log{\frac{p^2
e^\gamma}{\mu^2}}\right]+\textrm{IR terms}\bigg\}\nn\\
\Pi(\glug,\gluG;g)&=&p^2\bigg\{\left[\begin{array}{c}-67/9\\-64/9\end{array}\right]+\frac{11}{3}\left[-\frac{1}{\epsilon}+\log{\frac{p^2
e^\gamma}{\mu^2}}\right]+\textrm{IR terms}\bigg\}\nn\\
\Pi(+,+;g)&=&p^{+2}\left\{\left[\begin{array}{c}5/9\\8/9\end{array}\right]-\frac{1}{3}\left[-\frac{1}{\epsilon}+\log{\frac{p^2
e^\gamma}{\mu^2}}\right]\right\}\nn\\
\Pi(g^{\mu},g^{\nu};s)&=&-(p^2g^{\mu\nu}-p^\mu
p^\nu)\left\{\left[\begin{array}{c}4/9\\4/9\end{array}\right]-\frac{1}{6}\left[-\frac{1}{\epsilon}+\log{\frac{p^2
e^\gamma}{\mu^2}}\right]\right\}\nn\\
\Pi(g^{\mu},g^{\nu};q)&=&-(p^2g^{\mu\nu}-p^\mu
p^\nu)\left\{\left[\begin{array}{c}20/9\\20/9\end{array}\right]-\frac{4}{3}\left[-\frac{1}{\epsilon}+\log{\frac{p^2
e^\gamma}{\mu^2}}\right]\right\}\nn\\
\Pi(\fr{},\Fr{})&=&p^2\bigg\{\left[\begin{array}{c}-7\\-6\end{array}\right]+
3\left[-\frac{1}{\epsilon}+\log{\frac{p^2
e^\gamma}{\mu^2}}\right]+\textrm{IR
terms}\bigg\}\label{ALL_sfmass}\eea the IR terms for the self-mass
are simple enough to be listed\bea
\textrm{IR}_{self-mass}=\int_0^{p^+}{dq^+\left[\frac{-2}{p^+-q^+}+\frac{-2}{q^+}\right]\left[-\frac{1}{\epsilon}+\log{\frac{(p^+-q^+)q^+p^2
e^\gamma}{\mu^2p^{+2}}}\right]}\eea We observe that the quantity
$\Pi(g^I,g^J;g)+\Pi(g^I,g^J;q)/2-\Pi(q,\bar{q})$ is zero with
$\delta$-regulator or \dred, but non-zero with {\dreg}.

We also give the superficially divergent parts. To explain the
notations, $D$ is the dimension in which tensor algebra or gamma
matrix algebra is performed while $D_1$ is the dimension in which
momentum integral is done. So by taking $D_1=D=2-2\epsilon$ we
obtain the {\dreg} result, and by taking $D=2,\;D_1=2-2\epsilon$
we obtain the {\dred} result. The indices $I,\;J...$ range from
$1$ to $2$, and $i,j...$ range from $1$ to $2-2\epsilon$. For
{\dreg} we of course identify $I,J$ with $i,j$.\bea
\Pi(s,s)&=&p^2\frac{1}{\epsilon}(-4)\nn\\
\Pi(g^I,g^J;g)&=&p^2\bigg[\frac{1}{\epsilon}\left(-4\delta^{IJ}+\frac{D}{3D_1}\delta^{ij}\right)\nn\\\Pi(+,+;g)&=&p^{+2}\frac{1}{\epsilon}\frac{D}{6}+\textrm{Art.
Div.}\nn\\
\Pi(g^I,g^J;s)&=&p^2\delta^{ij}\frac{1}{\epsilon}\frac{1}{3D_1}\nn\\
\Pi(+,+;s)&=&p^{+2}\frac{1}{\epsilon}\frac{1}{6}\nn\\
\Pi(g^I,g^J;q)&=&p^2\frac{1}{\epsilon}\left(-\frac{4}{3D_1}\delta^{ij}+2\delta^{IJ}\right)\nn\\
\Pi(+,+;q)&=&p^{+2}\frac{1}{\epsilon}\frac{4}{3}\nn\\
\Pi(q,\bar{q})&=&-\frac{D}{2}\left[p_{\perp}^i\gamma^i-\frac{p_{\perp}^2}{2p^+}\gamma^+-{p^+}\gamma^-\right]+\frac{(D-8)p^2}{4p^+}\gamma^++\textrm{Art.
Div.}\nn\\
\label{ALL_sfmass_D}\eea The results above all contain a factor of
${ig^2N_c}/{(8\pi^2)}\trace[t^at^b]$.

\noindent\textbf{Triangle Diagrams}

We omit factor of $ig^3N_c/(8\pi^2)\Tr[t^at^bt^c]$ \bea
\Gamma(\glug,\glug,\gluG;s)&=&\frac{-2p_3^+}{p_1^+p_2^+}K^\wedge_{2,1}\left\{\left[\begin{array}{c}-{4}/{9}\\-{4}/{9}\end{array}\right]+\frac{1}{6}\left[-\frac{1}{\epsilon}+\log{\frac{p_o^2
e^{\gamma}}{\mu^2}}\right]-\alpha\frac{1}{6}\frac{p_1^+p_2^+}{p_3^{+2}}\right\}\nn\\
\Gamma(\glug,\glug,\gluG;q)&=&\frac{-2p_3^+}{p_1^+p_2^+}K^\wedge_{2,1}\left\{\left[\begin{array}{c}-{20}/{9}\\-{20}/{9}\end{array}\right]+\frac{4}{3}\left[-\frac{1}{\epsilon}+\log{\frac{p_o^2
e^{\gamma}}{\mu^2}}\right]+\alpha\frac{2}{3}\frac{p_1^+p_2^+}{p_3^{+2}}\right\}\nn\\
\Gamma(\glug,\glug,\gluG;g)&=&\frac{-2p_3^+}{p_1^+p_2^+}K^\wedge_{2,1}\left\{\left[\begin{array}{c}{67}/{9}\\{64}/{9}\end{array}\right]-\frac{11}{3}\left[-\frac{1}{\epsilon}+\log{\frac{p_o^2\delta
e^{\gamma}}{\mu^2}}\right]-\alpha\frac{1}{3}\frac{p_1^+p_2^+}{p_3^{+2}}\right\}+\textrm{IR
terms}\nn\\
\Gamma(\fr{},\Fr{},\gluG)&=&\frac{2p_2^+}{q^+p_1}K^\vee_{p_1,q}\left\{\left[\begin{array}{c}5\\6\end{array}\right]-3\left[-\frac{1}{\epsilon}+\log{\frac{p_o^2
e^{\gamma}}{\mu^2}}\right]\right\}+\textrm{IR terms}\nn\\
\Gamma(s,s,\glug)&=&\frac{2}{q^+}K^\wedge_{2,1}\left\{\left[\begin{array}{c}8\\8\end{array}\right]-4\left[-\frac{1}{\epsilon}+\log{\frac{p_o^2
e^{\gamma}}{\mu^2}}\right]\right\}+\textrm{IR
terms}\label{vert_loop}\eea where $\alpha=1$ if leg 3 is the
off-shell leg $p_o$, and zero otherwise. It can be checked that
the quantity
$\Gamma(g,g,g;g)+\Gamma(g,g,g;q)/2-\Gamma(q,\bar{q},g)$ is zero
only for $\delta$-regulator or {\dred}.

Next, we give the superficially divergent parts. The external
gluon self-mass
diagrams have been included in this case to be able to get an artificial divergence free expression. We omit $-ig^3N_c/(8\pi^2)\Tr[t^at^bt^c]$\bea%
\Gamma(g^I,g^J,g^K;g)&=&\left[\frac{1}{\epsilon}\left(-\frac{1}{3}-\frac{D}{6}+\frac{13D_1}{6}\right)+\frac{14}{3}\right]\delta^{IJ}\frac{1}{p_3^+}K^{k}_{ji}+\textrm{cyc. perm}\nn\\
\Gamma(g^I,g^J,g^K;f)
&=&\left[\frac{1}{\epsilon}(6-D_1)-\frac{10}{3}\right]\delta^{IJ}\frac{1}{p_3^+}K^{k}_{ji}+\frac{1}{\epsilon}\left(-\frac{8}{3}+\frac{2D_1}{3}\right)\delta^{ij}\frac{1}{p_3^+}K^{k}_{ji}+\textrm{cyc. perm}\nn\\
\Gamma(g^I,g^J,g^K;s)&=&-\left[\frac{1}{3}\delta^{IJ}\frac{1}{p_3^+}K^{k}_{ji}+\frac{1}{\epsilon}\left(\frac{2}{3}-\frac{D_1}{6}\right)\delta^{ij}\frac{1}{p_3^+}K^{k}_{ji}+\textrm{cyc.
perm}\right]\nn\\
\Gamma(s,s,g^K)&=&\left[\frac{1}{\epsilon}\left(-\frac{1}{3}-\frac{D}{6}+\frac{13D_1}{6}\right)+5\right]
\frac{1}{p_3^+}K^{k}_{ji}\nn\\
\Gamma(q,\bar{q},g^K)&=&\left[\frac{1}{\epsilon}\left(\frac{1}{3}-\frac{D}{12}+\frac{5D_1}{6}\right)+\frac{3}{2}\right]\gamma^+\frac{q^i}{q^+}\nn\\
&+&\left[\frac{1}{\epsilon}\left(-\frac{1}{2}-\frac{3D_1}{4}\right)-\frac{3}{2}\right]\gamma^K+\frac{1}{\epsilon}\left(\frac{1}{6}+\frac{D-D_1}{12}\right)\gamma^k\nn\\
&+&\left[\frac{1}{\epsilon}\left(-\frac{3}{4}+\frac{D}{8}+\frac{3D_1}{8}\right)+\frac{3}{4}\right]\left(\gamma^K\gamma^+\frac{\slashed{p_1}}{p_1^+}-\frac{\slashed{p_2}}{p_2^+}\gamma^K\gamma^+\right)\bigg]\eea
In this list, only the terms containing $\slashed{p_1}$ or
$\slashed{p_2}$ in $\Gamma(q,\bar{q},g^K)$ have artificial
divergence.

%\fi
\end{document}